\documentclass[a4, useAMS,usenatbib]{mn2e}

\pdfoutput=1
\usepackage[pdftex]{graphicx}
\usepackage[pdftex]{color}
\definecolor{linkblue}{rgb}{0,0,0.8}
\definecolor{linkgreen}{rgb}{0,0.5,0}
\usepackage[colorlinks,bookmarks,linkcolor=linkblue, citecolor=linkgreen, urlcolor=linkblue]{hyperref}
\hypersetup{linkcolor=linkblue, citecolor=linkgreen, urlcolor=linkblue}

\usepackage[T1]{fontenc}
\usepackage[utf8]{inputenc}
\usepackage{lmodern}
\usepackage[english]{babel}
\usepackage{amsmath, amssymb}

\providecommand{\adsurl}[1]{\href{#1}{ADS}}

\newcommand{\hloc}{H_{\text{loc}}}
\newcommand{\hbkg}{H_0}

\title[Uncertainty on $w$ from large-scale structure]{Uncertainty on $w$ from large-scale structure}

\author[Marra, P\"a\"akk\"onen and Valkenburg]{Valerio Marra$^{1}$, Mikko P\"a\"akk\"onen$^{2,3}$ and Wessel Valkenburg$^{1,4}$\\
$^{1}$Institut für Theoretische Physik, Universität Heidelberg, Philosophenweg 16, 69120 Heidelberg, Germany\\
$^{2}$Department of Physics, PL 35, 40014 University of Jyv\"askyl\"a, Finland\\
$^{3}$Helsinki Institute of Physics, PL 64, 00014 University of Helsinki, Finland\\
$^{4}$Instituut-Lorentz for Theoretical Physics, Universiteit Leiden, Postbus 9506, 2333 CA Leiden, The Netherlands}

\begin{document}

\date{Accepted XXX. Received XXX; in original form XXX}

\pagerange{\pageref{firstpage}--\pageref{lastpage}} \pubyear{2012}

\maketitle

\label{firstpage}

\begin{abstract}
We find that if we live at the center of an inhomogeneity with total density contrast $|\delta_0| \simeq 0.1-0.15$, dark energy is {\em not} a cosmological constant at 95\% confidence level.
Observational constraints on the equation of state of dark energy, $w$, depend strongly on the local matter density around the observer.
We model the local inhomogeneity with an exact spherically symmetric solution which features a pressureless matter component and a dark-energy fluid with constant equation of state and negligible sound speed, that reaches a homogeneous solution at finite radius.
We fit this model to observations of the local expansion rate, distant supernovae and the cosmic microwave background.
We conclude that the possible uncertainty from large-scale structure has to be taken into account if one wants to progress towards not just precision but also accurate cosmology.
\end{abstract}


\section{Introduction}

We have entered the so-called era of accurate cosmology \citep{Peebles:2002iq}.
The aim is to understand the composition and expansion history of the universe at the percent level. In particular, one of the goals is to establish if the observed acceleration of the universe \citep{Perlmutter:1998np, Riess:1998cb} is driven by the cosmological constant or by a fluid with negative pressure, dark energy.
At the most basic level the question is if the observed equation of state $w$ is compatible with $-1$ or not, the value corresponding to the cosmological constant.
It is therefore crucial to study all possible systematic effects on $w$ \citep{Amendola:2010ub,Sinclair:2010sb, Marra:2010pg, deLavallaz:2011tj, Romano:2011mx}.

As our observations are confined to the light cone, there is an intrinsic degeneracy between temporal evolution and spatial variation around us.
In particular, inhomogeneities around us are degenerate with the properties of dark energy, most importantly its equation of state.
A clear example of how intertwined are attempts to detect any evolution of dark energy to large-scale structures is given by the so-called ``void models''.
An observer inside a spherical underdensity expanding faster than the background sees indeed apparent acceleration, thus removing the need for dark energy \cite[see e.g.][and references therein]{Marra:2011ct}.
Void models strongly violate the Copernican principle and have been ruled out -- at least in their simplest incarnation -- as they predict a too strong kinematic Sunyaev-Zel'Dovich effect \citep{GarciaBellido:2008gd,Zhang:2010fa,Moss:2011ze,Zibin:2011ma}.
While on one hand this strengthens the case for dark energy as the likely explanation for the acceleration of the universe, on the other hand it illustrates how large-scale structure can alter the determination of cosmological parameters.
Therefore, it is necessary to adequately model large-scale structures if one has to achieve the grand goal of accurately determining the composition of the universe.

In \citet{Valkenburg:2011ty} it was shown by means of mock data that a local inhomogeneity, of proportions similar to a structure on the surface of last scattering that could cause the CMB Cold Spot, can have strong effects on our perception of the equation of state of dark energy. Here we extend that analysis to real data using the model of \citet{Marra:2011zp}.
More precisely, we consider a $w$CDM model endowed with a local almost-linear inhomogeneity surrounding the observer and test it against supernova observations, CMB anisotropies and local measurements of the Hubble parameter. By $w$CDM we mean a universe containing dark matter and a dark-energy fluid with equation of state $w$.
In this way we can show how the observed large-scale structure of the universe could impact the reconstruction of the dark-energy parameters.

Following \citet{Valkenburg:2011ty}, we consider an inhomogeneity inspired by the observed Cold Spot in the CMB, which has a radius of roughly $5^{\circ}$ and a temperature deviation of roughly $\mathcal{O}(50\sim200)$ $\mu$K.\footnote{See e.g.~\cite{Cruz:2006sv,Zhang:2009qg,Bennett:2010jb}.}
The idea is that the Cold Spot is a primary CMB anisotropy due to an object on the surface of last scattering, and not a secondary effect caused by an object along the line of sight \citep{Tomita:2005nu,Inoue:2006rd,Inoue:2006fn,Masina:2008zv}.
Such an inhomogeneity has a radius of roughly 1 Gpc and a density contrast today of roughly $-0.1$ \citep{Valkenburg:2011ty}.
It is therefore too shallow to bias the $w$CDM model to the point of removing the need for dark energy, as in the void scenario.
Nonetheless, as argued above, such a structure may bias the value of the dark-energy parameters to a level that may be important if one wants to determine whether dark energy is a cosmological constant or not.
Moreover, as we briefly argue in the body of this paper, structures of radius $\sim$1 Gpc and density contrast today $\sim$0.01 -- which still give an interesting effect -- are not at more than three times the dispersion of the density perturbations arising from a close to scale-invariant primordial spectrum.
Therefore, the setup considered in this paper is not in conflict with standard cosmology and, in particular, the Copernican principle~\citep{Valkenburg:2012td}.

We model the inhomogeneity with a particular case of the spherically symmetric solution presented in \citet{Marra:2011zp}, which features a pressureless matter component and a dark-energy fluid with constant equation of state and negligible sound speed $c_s$.
The possibility of a dark-energy fluid with negligible sound speed has been investigated in the literature under various assumptions. This generally requires a non canonical scalar field like $k$-essence and kinetic gravity braiding, as opposed to standard quintessence models with canonical scalar fields which always have $c_{s}=1$ \citep[see e.g.][and references therein]{Creminelli:2008wc, Bertacca:2008uf, Lim:2010yk,Bertacca:2010ct,Deffayet:2010qz,Li:2011sd}.
Here we choose this particular model because it significantly simplifies the dynamical equations and the numerical analysis as there are no pressure gradients that can generate peculiar velocities from an initially comoving motion.
We use this model phenomenologically so as to minimally extend the parameter space of the $w$CDM model by adding only two extra parameters: the radius of the inhomogeneity and its overall contrast. All the other initial conditions follow indeed rigidly.

The paper is organized as follows.
In Section \ref{model} we go briefly through the formalism of the model and its initial conditions, and in Section \ref{analysis} we explain how the cosmological data analysis has been performed.
We show in Section \ref{results} that the effect of local inhomogeneity on the dark-energy parameters can be important and that the inhomogeneity is not unlikely to occur. We  conclude in Section \ref{conclusions}.

\section{The model} \label{model}

We consider the case of an observer located at the center of an inhomogeneous sphere embedded in a flat $w$CDM universe.
The inhomogeneities are given by a pressureless matter component and by a dark-energy fluid with constant equation of state $w_{\rm out}$ and negligible sound speed (the subscript ``out'' refers to values at $r>r_{b}$ where $r_{b}$ is the comoving radius of the inhomogeneity). To be more precise, the sound horizon is much smaller than the inhomogeneity scale considered so that we can set $c_s=0$ throughout the paper.
In terms of the radius dependent non-adiabatic equation of state this means that we consider:
\begin{equation}
w(r,t) = w_{\rm out} \,   {\rho_{X, {\rm out}} (t)  \over   \rho_{X}(r,t)}  \,,
\end{equation}
that is, the pressure is homogenous:
\begin{equation}
p_{X} (r,t)=p_{X, {\rm out}} (t) = w_{\rm out}  \, \rho_{X, {\rm out}} (t) \,,
\end{equation}
where $r$ is the coordinate radius, $t$ is cosmic time, $\rho$ denotes energy density, $p$ denotes pressure and the label $X$ refers to the dark-energy fluid.
Since pressure gradients are absent, matter and dark energy evolve along geodesics.
Moreover, we set initial conditions such that dust and dark energy are initially comoving. Therefore, the absence of pressure gradients implies that peculiar velocities between the two fluids will never develop and that the matter and dark-energy reference frames always coincide.
Next, we discuss the equations governing the dynamics of the model and the relevant initial conditions.

\subsection{Dynamical equations} \label{equa}

We adopt the exact spherically symmetric inhomogeneous solution with $n$ perfect fluids presented in \citet{Marra:2011zp}, which we limit to the case discussed above.
We will now briefly introduce the relevant equations, and we refer to \citet{Marra:2011zp} for the general equations and more details.

Using the reference frame of the dust and dark-energy components, the metric describing our model is:
\begin{equation} \label{metric}
   \textrm{d}s^2=-\textrm{d}t^2+\frac{Y'(r,t)^2}{1-k(r) r^2}\textrm{d}r^2+Y(r,t)^2\textrm{d}\Omega^{2} \,, 
\end{equation}
where $Y(r,t)$ is the scale function, $k(r)$ is the curvature function, $\textrm{d}\Omega^{2}= \textrm{d} \theta^{2} + \sin^{2}\theta \, \textrm{d} \phi^{2}$ and we have set $c=1$.
A prime denotes partial derivation with respect to the coordinate radius $r$, whereas a dot denotes partial derivation with respect to the coordinate time $t$.
The curvature function is time independent because of the adopted reference frame and sound speed~\citep{Marra:2011zp}.
The metric (\ref{metric}) is written in the same form as the Lema\^itre-Tolman-Bondi (LTB) metric. However, with the inclusion of a dark-energy fluid, the dynamics is no longer that of the LTB metric.
The metric of Eq.~(\ref{metric}) reduces to the Friedman-Lema\^itre-Robertson-Walker (FLRW) metric if $k(r) =$const and $Y(r,t)=r \, a(t) $, where $a(t)$ is the scale factor.

We  label the dust component with $M$ and the dark-energy component with $X$.
The conservation equation for the dust source can be solved directly and gives:
\begin{equation} \label{dustevo}
{\rho_{M}(r, t) \over \rho_{M}(r, \bar t)}   =    {Y^{2}(r, \bar t) Y'(r, \bar t)  \over  Y^{2}(r, t) Y'(r, t) }   \,,
\end{equation}
where $\bar t$ is the initial time at which we give the initial conditions.
As explained before, there are no peculiar velocities between the two fluids and the remaining dynamical equations reduce to:
\begin{align}
\dot Y^2(r,t) =&  \frac{2 G F(r,t)}{Y(r,t)} -k(r) r^2  ,  \label{set1} \\
\dot Y'(r,t) =&  \frac{G F'(r,t)}{Y(r,t) \dot Y(r,t)} -\frac{G F(r,t) Y'(r,t)}{Y^2(r,t) \dot Y(r,t)} - { [k(r)r^2]' \over 2 \dot Y(r,t)}  ,  \label{set2} \\
\dot F(r,t) =&- 4 \pi Y^{2}(r,t) \dot Y(r,t)  \; p_{X, {\rm out}} (t)   ,   \label{set3}\\
\dot \rho_{X}(r,t)  =&  - \Big [\rho_{X}(r,t)+p_{X, {\rm out}} (t) \Big ] \Big [H_R(r,t) + 2 H_A(r,t) \Big]  ,    \label{set4}
\end{align}
where the radial and angular expansion rates are $H_R = \dot{Y}' /Y'$ and $H_A=\dot{Y} /Y$, and $F$ is the total effective gravitating mass which also satisfies the following consistency equation:
\begin{equation}
F'(r,t) = 4 \pi Y^{2}(r,t) Y'(r,t)  \Big [ \rho_M (r,t) + \rho_{X}(r,t) \Big ] \,.
\end{equation}
Eq.~(\ref{set2}) is the $r$-derivative of Eq.~(\ref{set1}) and allows us to solve directly for the unknown functions $Y$, $Y'$, $F$ and $\rho_X$ without having to take numerical derivatives.
If $w=w_{\rm out}=-1$, this solution becomes the usual $\Lambda$LTB model which has been studied recently in, e.g.,~\citet{Enqvist:2006cg, Sinclair:2010sb, Marra:2010pg, Valkenburg:2011tm, Romano:2011mx}.

Finally, in this particular case the light-cone equations have the same form as in the LTB model:
\begin{equation}
            \frac{dt}{dz} = -\frac{Y'}{(1+z)\dot{Y}'} \,, \qquad
            \frac{dr}{dz} =\frac{\sqrt{1-k(r) r^2}}{(1+z)\dot{Y}'}   \,.
\end{equation}

\subsection{Initial and boundary conditions} \label{inico}

\begin{figure}
\begin{center}
\includegraphics[width= \columnwidth]{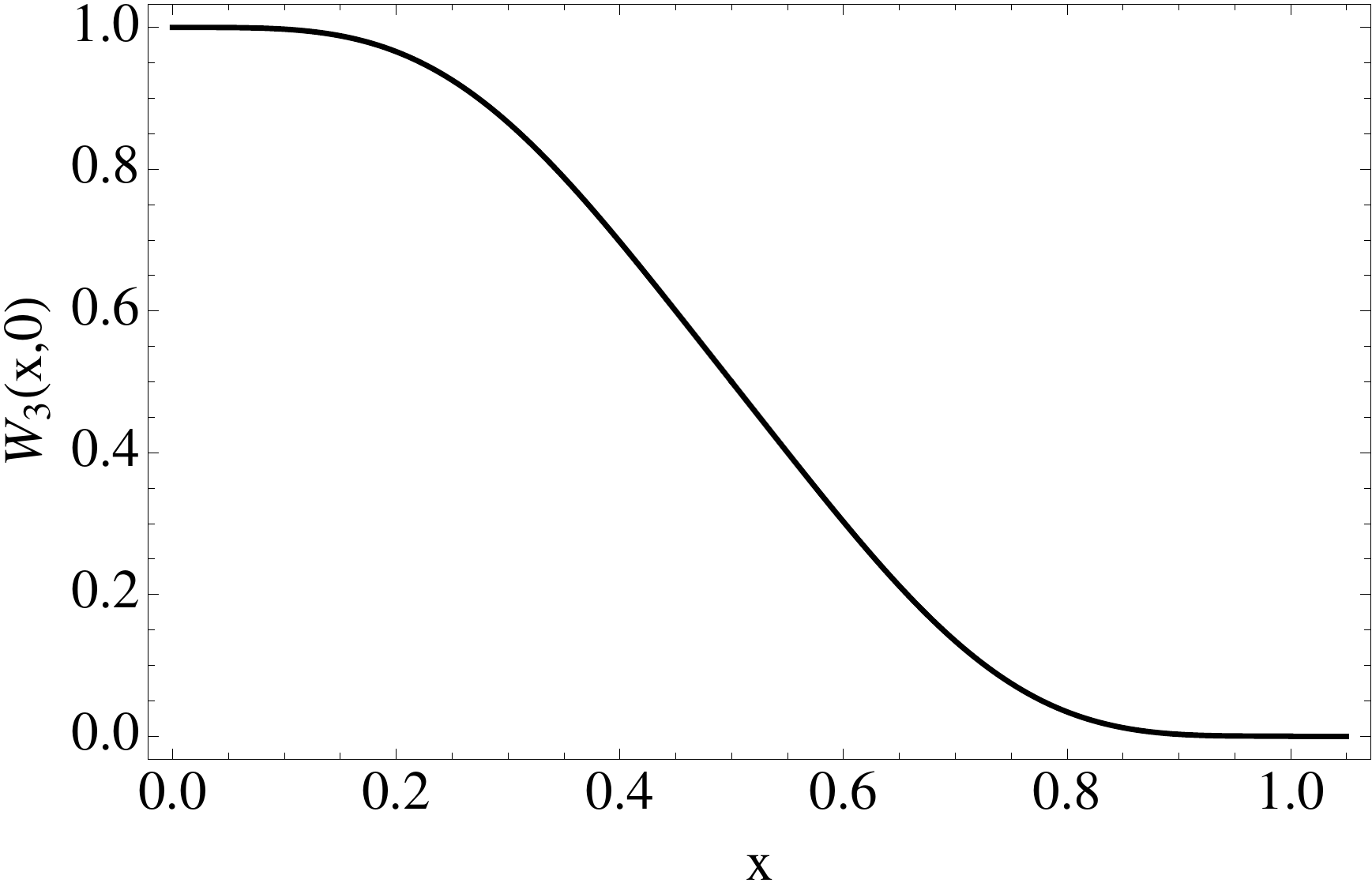}
\caption{Shape of the auxiliary function $W_3\left(x,0\right)$ that is used in Eq.~(\ref{profi}) to model the curvature profile.}
\label{fig:profile}
\end{center}
\end{figure}

We fix the flat $w$CDM background model by setting $h$, $\Omega_{X}$ and $w_{\rm out}$, where $\Omega_{X}$ is the present-day background dark-energy density parameter and $h$ is the present-day dimensionless Hubble rate defined by $\hbkg=100 \, h$ km s$^{-1}$ Mpc$^{-1}$.

We fix the gauge for the radial coordinate in Eq.~\eqref{metric} such that \mbox{$Y(r, \bar t)= r \, a(\bar t)$} at some initial time $\bar t$.
We choose to parametrize the curvature by means of
\begin{align} \label{profi}
k(r) =& k_c \; W_3\left(\frac{r}{r_b},0\right) \,,
\end{align}
where $r_b$ is the comoving radius of the spherical inhomogeneity and:
\begin{align} \label{eq:theprofile}
W_3\left(x,0\right)=\left\{\begin{array}{ll}
\frac{1}{4\pi^2}  + 1 - 2 x^2 -  {\cos \left( 4\pi x \right) \over 4\pi^2} & \mbox{for }  0 \le x < \frac{1}{2}\\
 \frac{-1}{4\pi^2}  + 2 \left (1 -x \right)^2 +  {\cos \left( 4\pi x \right) \over 4\pi^2} & \mbox{for }  \frac{1}{2}  \le x < 1,\\
0& \mbox{for } x\geq 1
\end{array}\right. 
\end{align}
is the third order of the function $W_n(x,\alpha)$, which has been defined in~\citet{Valkenburg:2011tm} and interpolates from $1$ to $0$ in the interval $\alpha < x < 1$ while remaining $C^n$ everywhere.
Hence $k(r)$ is $C^3$ everywhere, such that the metric is $C^2$ and the Riemann curvature is $C^0$.
Although the function $W_3\left(x,0\right)$ looks rather complicated, it has actually a very simple shape as one can see in Fig.~\ref{fig:profile}.
The curvature profile of Eq.~(\ref{profi}) is exactly zero for $r \ge r_b$ and so the metric is correctly matched to the exterior spatially-flat $w$CDM model, which means that the central over- or under-density is automatically compensated by a surrounding under- or over-dense shell.
The constant $k_c$ gives the curvature at the center of the inhomogeneity and will determine its density contrast.
A spherical inhomogeneity depends crucially on only two {\it physical} parameters: the radius and the overall density contrast.
Therefore, the precise shape of the density profile should not be essential and our analysis should be representative also of other possible curvature or density profiles.

Next, we have to give initial conditions at $t=\bar t$ for $F(r, \bar t)=\bar F_{M}(r)+\bar F_{X}(r)$.
We choose $\bar t$ such that matter is dominant over dark energy, $\bar F_{X}(r) \ll \bar F_{M}(r)$, and so the model becomes the standard dust LTB model.
We can then link the curvature function $k(r)$ to the initial condition for $\bar F_{M}(r)$ by demanding that the universe has the same age $\bar t$ for any $r$. That is, we demand a homogeneous Big Bang, implying the absence of decaying modes in the matter density \citep{Zibin:2008vj}.
In particular we can use the following analytic result of \citet{VanAcoleyen:2008cy} valid for a linear matter density contrast:
\begin{equation}
k(r) \simeq  {5\over 3}  a^2(\bar t)  H_{\rm out}(\bar t) ^2  \; \bar \delta_{F_M}(r) \,,
\end{equation}
which clearly relates the curvature at the center $k_c$ to the matter contrast at the center $\bar \delta_{M} =\bar \delta_{F_M}(0)$.
The latter term is the contrast in the gravitating mass and is defined as:
\begin{equation}
\bar \delta_{F_M}(r) = {\bar F_{M}(r) \over \bar F_{M, {\rm out}}(r)} -1 \,,
\end{equation}
where $\bar F_{M, {\rm out}}= {4 \pi \over 3}  a^{3}(\bar t) r^{3} \rho_{M, {\rm out}} (\bar t)$ is the corresponding background gravitating mass and the gauge $Y(r, \bar t)= r \, a(\bar t)$ has been used.
The initial matter density is then:
\begin{equation}
\rho_{M}(r,\bar t)= {\bar F_{M}'(r)  \over 4 \pi a^3(\bar t) r^2 } \,.
\end{equation}

In order to have initial conditions without decaying modes in the dark-energy component, we have to set its initial profile according to the following relation valid during matter domination and $c_{s}^{2}\ll1$ \citep{Ballesteros:2010ks}:
\begin{equation}
{\bar \delta_{F_X}(r)  \over \bar \delta_{F_M}(r) }= {\bar \delta_{X} \over \bar \delta_{M}} =  {1+w_{\rm out} \over 1-3 w_{\rm out}} \,,
\end{equation}
where, analogous to the matter contrast, we have,
\begin{equation}
\bar \delta_{F_X}(r) = {\bar F_{X}(r) \over \bar F_{X, {\rm out}}(r)} -1 \,,
\end{equation}
$\bar F_{X, {\rm out}}= {4 \pi \over 3}  a^{3}(\bar t) r^{3} \rho_{X, {\rm out}} (\bar t)$ and $\bar \delta_{X} =\bar \delta_{F_X}(0)$.
The initial dark-energy density is then,
\begin{equation}
\rho_{X}(r,\bar t)= {\bar F_{X}'(r)  \over 4 \pi a^3(\bar t) r^2 } \,.
\end{equation}
Hence, all the initial conditions relative to the inhomogeneous patch are indeed specified by a given curvature profile~$k(r)$.

\section{Cosmological data analysis} \label{analysis}

In this Section we  explain how we compare the predictions of this model with supernovae, Hubble rate and cosmic microwave background observations.
We decided not to include baryon acoustic oscillations in the analysis as perturbation theory in an inhomogeneous background has not been thoroughly understood yet\footnote{See, however, \citet{Nishikawa:2012we} for a recent development.} \citep{Zibin:2008vj,Clarkson:2009sc,Alonso:2010zv}.

\subsection{Hubble rate}\label{subsec:h0}

The Hubble rate is obtained by measuring cosmological standard candles mostly within a redshift range with median value $z_h \sim 0.05$.
We compare the observed value to the theoretical quantity,
\begin{equation} \label{hloco}
\hloc=  {1  \over z_{\rm max} - z_{\rm min}}  \int_{z_{\rm min}}^{z_{\rm max}}  H_A(r(z),t(z)) \,  dz \,.
\end{equation}
The values $z_{\rm max}$ and $z_{\rm min}$ depend on the redshift volume that is probed by a given experiment.
The reason we compare an averaged expansion rate to the data is primarily because the observed expansions rate in fact is an averaged quantity, so this should be a fair comparison. Moreover, when the redshift $z_b$ of the boundary of the inhomogeneity is close in value to~$z_h$, the averaged $\hloc$ may differ significantly from $H_A(r(z_h),t(z_h))$, which falsely would lead  to a bad fit.
The approach of Eq.~(\ref{hloco}) is to some extent arbitrary and one should instead reanalyze the raw data without assuming a FLRW fiducial model as it is usually done \citep[see e.g. the discussion in][]{Zumalacarregui:2012pq}. However, as we will see in Section \ref{scmb} our results depend weakly on the Hubble parameter constraint and so this caveat should not sizably affect our findings.

We  mainly consider the 
determination of the Hubble rate from \citealt{Riess:2009pu} (R09). However, in order to study a possible sensitive dependence on this datum, we will also consider the results from \citealt{Freedman:2000cf} (F01) and \citealt{Sandage:2006cv} (S06). The three measurements are:
\begin{align}
H_{\rm{F01}}&=72\phantom{.0} \pm 8\phantom{.0} \; \frac{\rm{km/s}}{\rm{Mpc}}, &0.005<z<0.1, \label{F01} \\
H_{\rm{S06}}&=62.3 \pm 6.3 \; \frac{\rm{km/s}}{\rm{Mpc}}, &0.01<z<0.07,   \label{S06} \\
H_{\rm{R09}}&=74.2 \pm 3.6 \; \frac{\rm{km/s}}{\rm{Mpc}}, &0.023<z<0.1.  \label{R09} 
\end{align}

\subsection{Supernova observations}

We use the Union2 SN Compilation \citep{Amanullah:2010vv}, which consists of 557 type Ia supernovae in the redshift range $z=0.015-1.4$.
As we are considering an almost-linear inhomogeneity surrounding the observer, we are not departing strongly from the standard model.
Therefore it should be a good approximation to use the magnitude-redshift and correlation tables provided by \citet{Amanullah:2010vv}.

\subsection{Cosmic microwave background} \label{scmb}

The metric of Eq.~(\ref{metric}) is matched to the background FLRW metric at a redshift at which radiation is still negligible.
In this way the last scattering surface, which is responsible for most of the CMB anisotropies, is outside the inhomogeneous patch and a standard analysis of the primordial CMB power spectrum is possible.
One has to replace the inhomogeneous model with an {\it effective} FLRW metric which accounts for the different angular diameter distance to the surface of last scattering of the CMB as compared to the homogeneous background model.
This is done by placing an FLRW observer ($\delta_0\equiv0$) in the same coordinate system at $r=0$ but at a different time than $t_0$, such that this observer's angular diameter distance to the surface of last scattering, which lies at some constant time $t_{\rm LS}$, agrees with the actual LTB observer's angular diameter distance to the surface of last scattering. The physics at last scattering itself is unaffected, since the {\em effective} FLRW observer is placed in the same FLRW universe in which the LTB patch is embedded, albeit at a different time. The CMB spectrum is then calculated using {\sc camb} \citep{Lewis:1999bs}.
See \citet{Biswas:2010xm, Moss:2010jx,Marra:2010pg} for more explicit details about how the effective model is obtained.
Note  that there are other contributions to the CMB coming from secondary effects, which in the inhomogeneity may differ from those in the effective FLRW metric, and are due to the photons traveling through  inhomogeneities inside the void, such as the late-time ISW effect and weak lensing. For the same reason as for which we ignore the baryon acoustic oscillations, we ignore these secondary effects, since they are subdominant and studying them would require knowledge of the growth of perturbations in an inhomogeneous background.
We  fit the theoretical predictions of our model to the WMAP 7-year data release \citep{Komatsu:2010fb}.

\subsection{Parameter estimation}

We perform a Markov-Chain Monte-Carlo likelihood analysis using {\sc CosmoMC} \citep{Lewis:2002ah}. We calculate all distance measures using an improved version of {\sc VoidDistancesII}{\footnote{\url{http://web.physik.rwth-aachen.de/download/valkenburg/}}} \citep{Biswas:2010xm}, which now acts as a wrapper around {\sc camb} \citep{Lewis:1999bs}, necessitating no changes to {\sc camb}'s source code and minimal changes to {\sc CosmoMC}'s source code. We combine this module with the $\Lambda$LTB module {\sc ColLambda}\footnote{\url{http://web.physik.rwth-aachen.de/download/valkenburg/ColLambda/}} \citep{Valkenburg:2011tm} for calculating all metric functions, which we extended to include the numerical solutions to the scenario discussed here, with $w_{\rm out}\neq -1$ and $c_s = 0$.
With this setup, for every selected vector of parameter values, we calculate the theoretical predictions for supernova distances, the local Hubble rate
and the CMB power spectrum. 
For reference, next to the inhomogeneous model we analyze its homogeneous background model, $w$CDM,  which is described by the same model but has $z_b\equiv0$ and $\delta_0\equiv0$.

\newlength{\mywidth}
\setlength{\mywidth}{.9 \columnwidth}
\begin{table}
\begin{center}
\begin{tabular}{c}
\begin{tabular*}{\mywidth}{c}
\hline
Flat priors\\
\hline \hline
\begin{tabular}{rcl}
$0.4\quad <$ &$h$ & $< \quad 1$\\
$0.005 \quad <$&$\Omega_{\rm b}h^2$ & $< \quad 0.1$ \\
$0.001\quad <$&$\Omega_{\rm dm}h^2$ & $< \quad 0.99$ \\
$-2\quad <$&$w_{\rm out}$ & $< \quad -0.4$ \\
$0.01 \quad <$ & $\tau$ & $ < \quad 0.8$ \\
$2.7 \quad <$ & $\log 10^{10} A_S$ & $ < \quad 4$ \\
$0.5 \quad <$ & $n_S$ & $ < \quad 1.5$ \\
$-0.2 \quad <$ & $\alpha_S$ & $ < \quad 0.2$ \\
$-0.2 \quad <$&$ \delta_0 $ & $< \quad 0.2$ \\
$100$~Mpc $\quad <$&$d(r_b) $ & $< \quad3$~Gpc
\end{tabular}\\\hline\\  \hline
\hline
Additional constraints\\
\hline \hline
\begin{tabular}{rl}
$\Omega_{X}$&$ >0$\\
$\Omega_{k}$&$ =0$
\end{tabular}\\
\hline
\end{tabular*}
\end{tabular}
\end{center}
\caption{Priors imposed on the parameters in the numerical analysis. The size of the LTB patch $d(r_b)$ is defined in Eq.~\eqref{eq:defdr}. The additional constraint $\Omega_X>0$ with $\Omega_k=0$, in fact implies a non-flat prior on both $\Omega_{\rm dm} h^2$ and $h$, as explained in Appendix~\ref{prioco}.}\label{tab:priors}
\end{table}

We take flat priors on the parameters listed in Table~\ref{tab:priors}. Most of these are the usual cosmological parameters: the background present-day Hubble rate $h$, the background baryon density $\Omega_{\rm b}h^2$, the background dark-matter density $\Omega_{\rm dm}h^2$, the equation of state of dark energy $w_{\rm out}$, the optical depth to re-ionization $\tau$, the amplitude of primordial scalar perturbations $A_S$, the tilt of the spectrum of primordial scalar perturbations $n_S$ and its running $\alpha_S$. We set the spatial curvature outside the LTB patch, $\Omega_k$, as well as the amplitude of primordial tensor perturbations to zero.

As discussed in Section~\ref{inico}, two additional parameters describe the LTB patch: the curvature at the center $k_c$ and the comoving radius $r_b$.
As shown in Table~\ref{tab:priors}, we will use as actual parameters the present-day total density contrast and the present-day proper size of the radius, respectively.
The latter is given by
\begin{align}
d(r_{b}) &\equiv \int_0^{r_{b}} \frac{Y'(r,t_0)}{\sqrt{1-k(r) r^2}} \,\textrm{d}r \,, \label{eq:defdr}
\end{align}
and we define the former as
\begin{align} \label{totcont}
\delta_0 \equiv \frac{\rho_{\rm in} - \rho_{\rm out}}{ \max(\rho_{\rm in},\rho_{\rm out})} \,,
\end{align}
where $\rho_{\rm in}= \rho_M (0,t_0) + \rho_X (0,t_0)$ and $\rho_{\rm out}=\rho_{M, {\rm out}} (t_0)+\rho_{X, {\rm out}} (t_0)$.
We calculate the contrast using the total density as in principle the dark-energy component can be as inhomogeneous as the matter component, and $\delta_0$ should be a relevant physical quantity to be used.
We chose a somewhat unusual definition by using $\max(\rho_{\rm in},\rho_{\rm out})$ in the denominator. With this definition, this quantity fundamentally satisfies $-1<\delta_0<1$, such that the parameter-space volume in over- and under-densities is equally distributed.
With the more usual definition we would have had $-1<\frac{\rho_{\rm in} - \rho_{\rm out}}{ \rho_{\rm out}}<\infty$, which in the Bayesian parameter estimation 
induces a strong prior favouring large over-densities, possibly excluding under-densities from the analysis. In practice, however, the preferred values for this parameter are small, such that the difference between both definitions is almost negligible.

We discuss the priors in more detail in Appendix~\ref{prioco}. In the next section, where we discuss the results of the MCMC parameter estimation, we explore the full parameter space, never fixing the background parameters to some central value. Therefore we can always marginalize over all parameters, and  do not bias the result in any way.
That is, any possible degeneracy, expected or unexpected, between the cosmological parameters and the LTB parameters will show up and will not influence the results without being noticed.

\section{Results} \label{results}

\begin{figure*}
\begin{center}
\includegraphics[width= .48\textwidth]{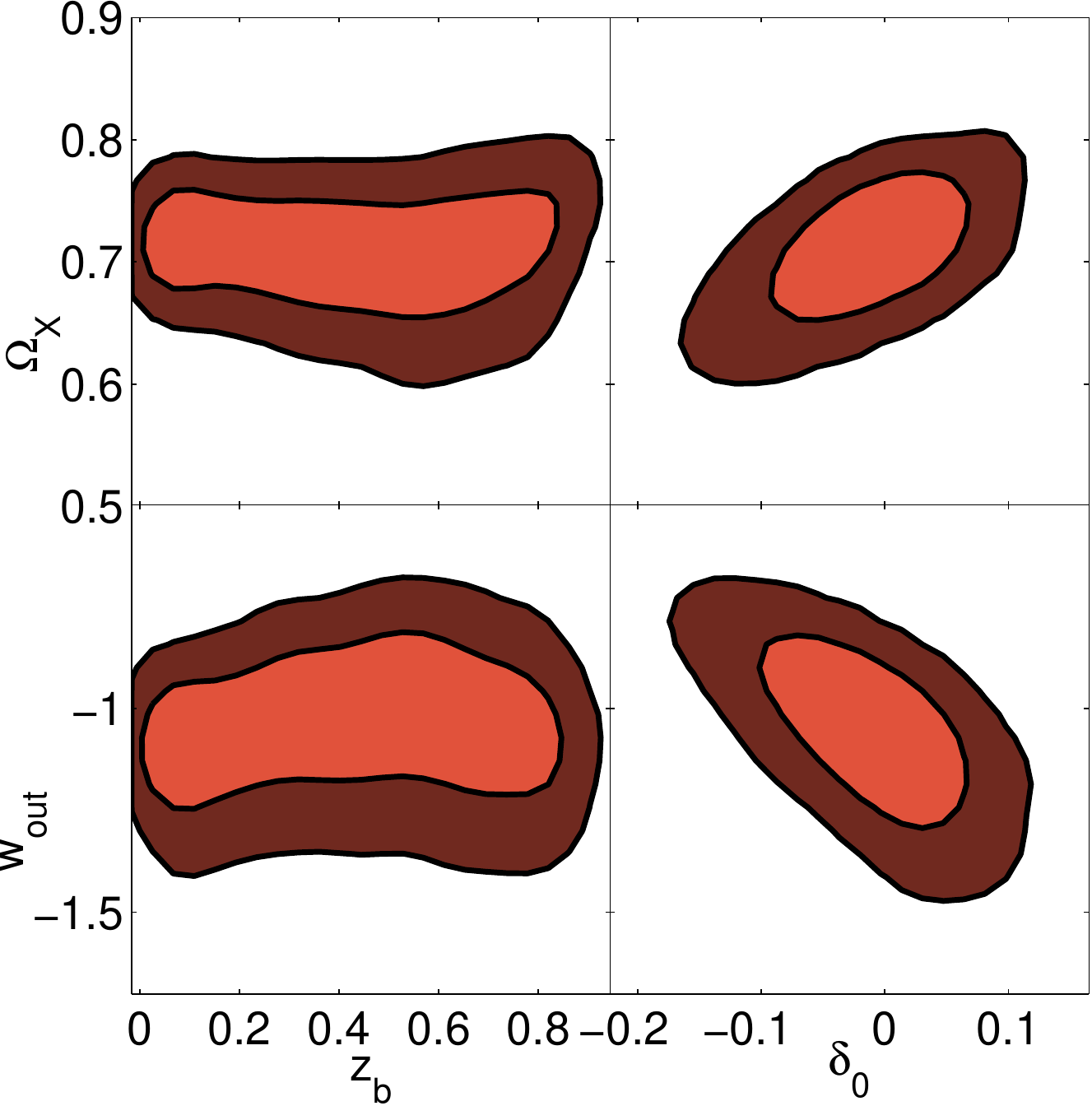}
    \qquad
\includegraphics[width= .48\textwidth]{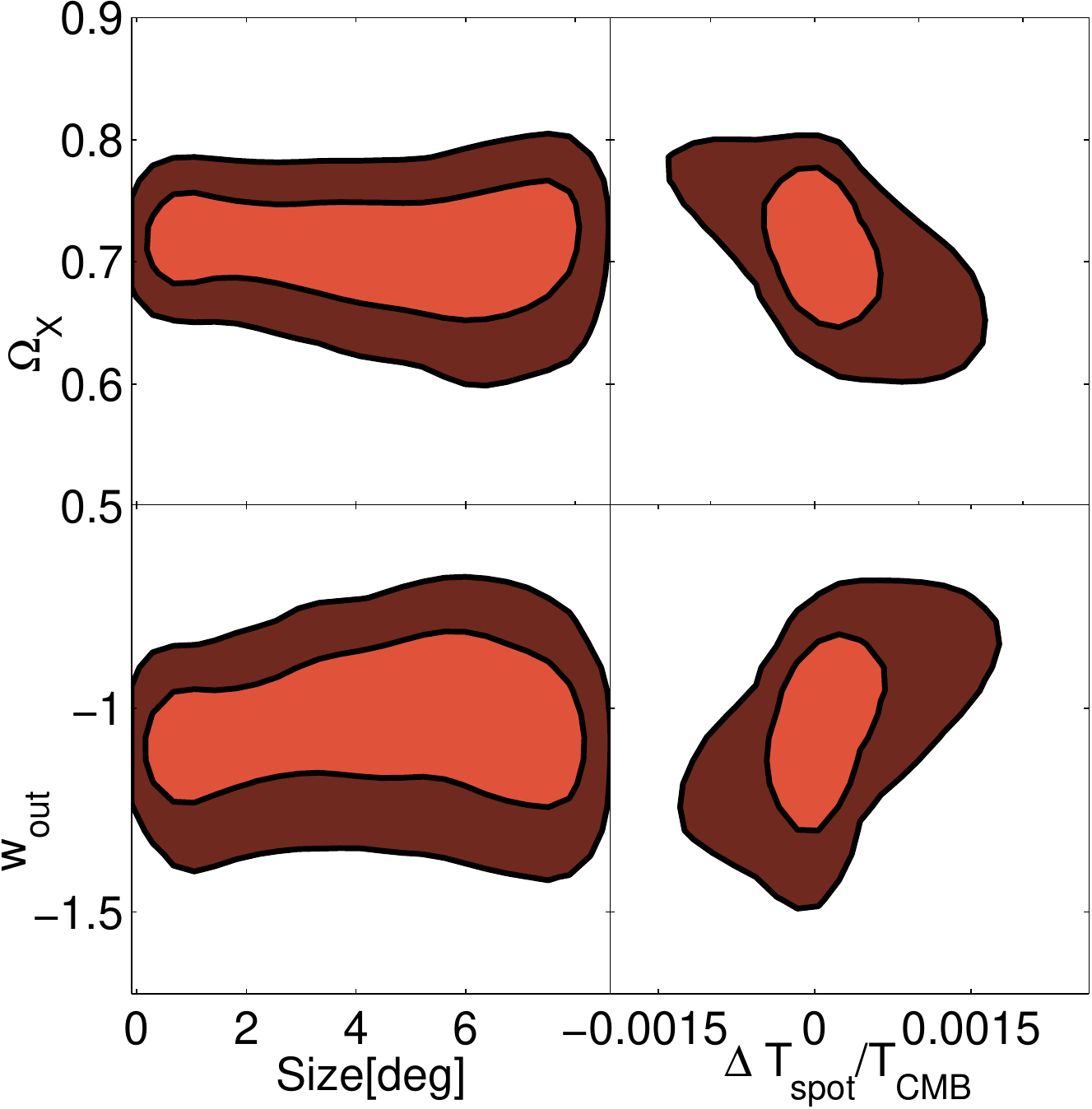}
\caption{Two-dimensional marginalized posterior probability distributions for the most interesting parameters characterizing the $w$CDM model endowed with a local inhomogeneity considered in this paper. The parameters are constrained by the Union2 SN Compilation \citep{Amanullah:2010vv}, the WMAP 7-year CMB power spectrum \citep{Komatsu:2010fb}
and the recent determination of the Hubble rate by \citet{Riess:2009pu}. Inner tangerine-tango coloured contours are 68\% confidence level (c.l.) contours, while the outer dark-red coloured contours are 95\% c.l. The left and right figures show the same information, however the redshift of the boundary of the inhomogeneity, $z_b$, and the total density contrast, $\delta_0$, on the left, are on the right traded in for the angular diameter that the inhomogeneity would have if it were located at the observer's last scattering surface and the temperature fluctuation that it would induce, respectively.}
\label{fig:2Dtango}
\end{center}
\end{figure*}

The four left panes in Figure~\ref{fig:2Dtango} show the parameters on which the main focus in this paper lies.
This figure shows the two-dimensional marginalized posterior probability distributions of the background dark-energy density $\Omega_X$, the dark-energy equation of state $w_{\rm out}$, the boundary of the inhomogeneity in redshift space $z_b$, and the total density contrast in the inhomogeneity $\delta_0$ (see Eq.~(\ref{totcont})).
Alternatively, in the four right panes of Figure~\ref{fig:2Dtango} we use as proxies for $z_b$ and $\delta_0$ the apparent size that the inhomogeneity would subtend if located at the last scattering surface and its corresponding temperature anisotropy, respectively. Here we define the temperature perturbation $\Delta T/T$ as the relative difference in CMB temperature at the center of the inhomogeneity and outside, in the homogeneous (average) background. This is not necessarily representative for the average temperature in the spot.

\begin{figure*}
\begin{center}
\includegraphics[width= .7\textwidth]{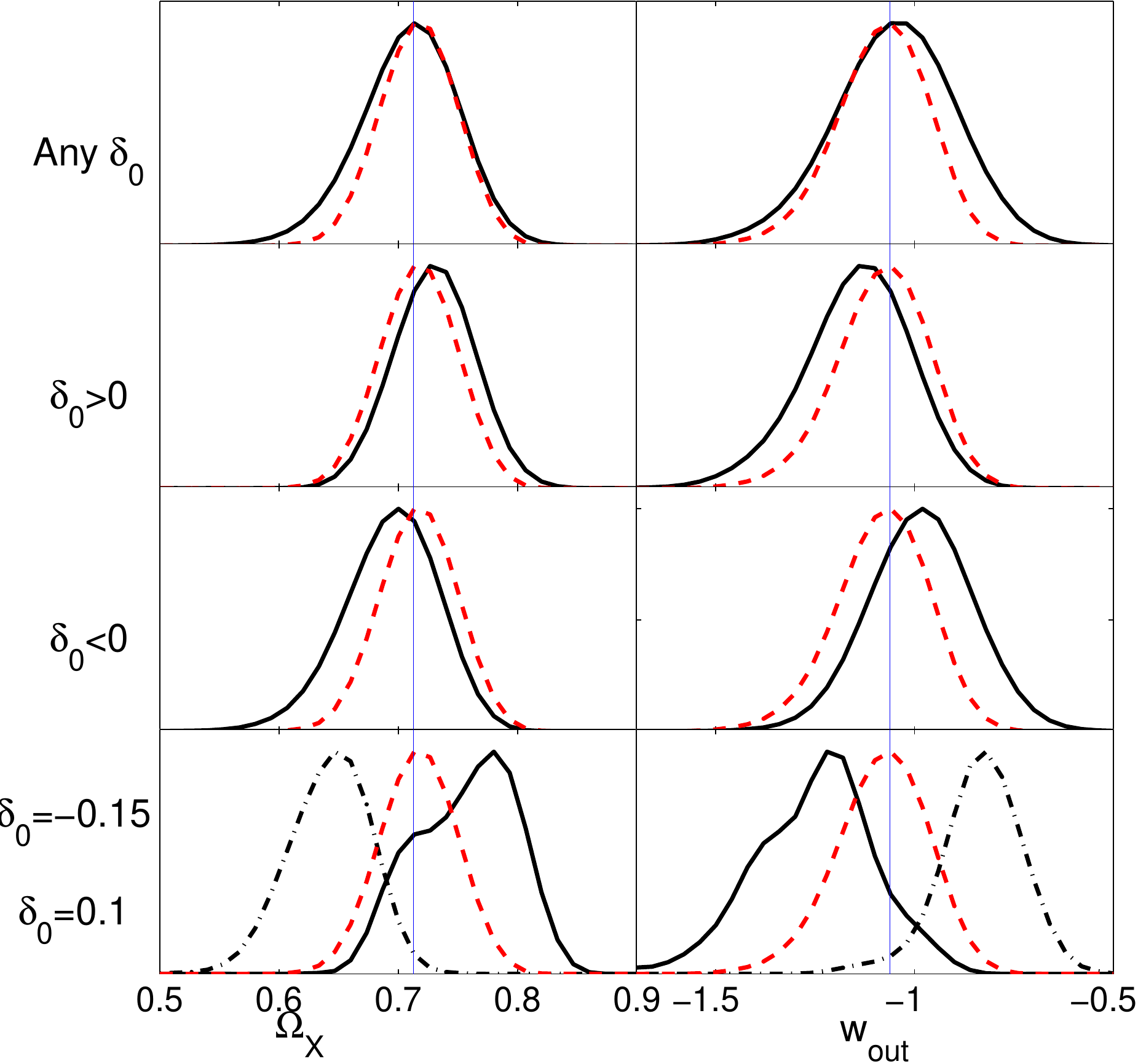}
\caption{One-dimensional marginalized posterior probabilities for $\Omega_X$ and $w_{\rm out}$, constrained by CMB, SNe, and $\hloc$.
The standard $w$CDM constraints on these parameters are shown in red dashed lines and are the same in the different panes. The blue vertical line serves as a guide for the eye, always going through the maximum of the $w$CDM value.
The constraints on the $w$CDM model endowed with a local inhomogeneity considered in this paper are given under different priors on $\delta_0$: $-0.2<\delta_0<0.2$ (top), $0<\delta_0<0.2$ (second row from top), $-0.2<\delta_0<0$ (third row from top), and for two constraining priors on $\delta_0$ (bottom): $\delta_0=0.1$ (solid black line) and $\delta_0=-0.15$ (dashed-dotted black line). A prior on $\delta_0$ that averages around zero, widens but does not shift the posterior distributions. If we know instead the sign of $\delta_0$, constraints on $w_{\rm out}$ and $\Omega_X$ shift by as much as $5\%\sim10\%$. In the bottom row, for $\delta_0=0.1$ the 95\% c.l.~upper bound on $w_{\rm out}$ is -1.03. For $\delta_0=0.15$ the 95\% c.l.~lower bound on $w_{\rm out}$ is -0.98. Both priors hence rule out the cosmological constant at 95\% c.l., given current observations. }
\label{fig:1Ddiffdelta}
\end{center}
\end{figure*}

The most interesting result is the clear degeneracy between $\delta_0$ and $w_{\rm out}$ in the lower right pane in the left of Figure~\ref{fig:2Dtango}, and 
the degeneracy between $\delta_0$ and $\Omega_X$ in the upper right pane. This graph explicates the necessity of properly modeling the inhomogeneity of the local universe, before any conclusion can be drawn on the properties of dark energy, in particular about the fundamental value of its equation of state: if the local density is ignored, the value of $w_{\rm out}$ can be misestimated by possibly 50\%. Note also that the redshift up to which the inhomogeneity extends, $z_b$, is hardly of any influence on the central value of $w_{\rm out}$ or $\Omega_X$.
The bias on the parameters is indeed of opposite sign for opposite $\delta_0$. Therefore, as we marginalize over $\delta_0$ in the combined posterior of $w_{\rm out}$ and $z_b$ (lower left pane in Fig.~\ref{fig:2Dtango}), the total effect is compensated (we will come back to this point with Figure~\ref{fig:1Ddiffdelta}). Note, however, that the scatter does increase with the size of the inhomogeneity.
The right of Figure~\ref{fig:2Dtango} shows the dimensions that the inhomogeneity would have on the observed CMB temperature map, if it were centered on the observer's surface of last scattering. In this situation there are hence two identical inhomogeneities in the universe: one surrounding the observer, one centered on the observer's surface of last scattering.
This figure shows that even spots that do not violate the observables of Section \ref{analysis}, do induce a strong bias on $w_{\rm out}$, following the findings of \citet{Valkenburg:2011ty}.
As before we see that the apparent size of the inhomogeneity has almost no effect on the value of $w_{\rm out}$ or~$\Omega_X$.

In Figure~\ref{fig:1Ddiffdelta} we show the one-dimensional marginalized posterior probabilities for $\Omega_X$ and $w_{\rm out}$, under different priors on $\delta_0$, imposed by means of importance sampling on the MCMC chains that explore the full range of $\delta_0$. The result is displayed in black, and for comparison the constraints on these parameters for the homogeneous $w$CDM model ($z_b\equiv 0$, $\delta_0\equiv0$) are displayed in dashed red lines. The top row shows $\Omega_X$ and $w_{\rm out}$ marginalized over all values of $\delta_0$, both positive and negative. Since positive and negative values of $\delta_0$ have opposing effects on $w_{\rm out}$ and $\Omega_X$, marginalizing over $\delta_0$ mostly widens the tails of the distributions.
This is already an important result showing how inhomogeneities contribute to the error budget in the cosmological parameters. This uncertainty from large-scale structures is expected to become more important when future data will tighten the confidence regions of the parameters of interest.
If we impose, however, the prior that $\delta_0>0$ (second row in Figure~\ref{fig:1Ddiffdelta}) or $\delta_0<0$ (third row from top), we find even stronger results: we see indeed that both the tails and the central values of $\Omega_X$ and $w_{\rm out}$ shift. This shift is significant if one wants to progress towards not just precision but also accurate cosmology. If we push the magnitude even further, pretending we know that the local density must be either $\delta_0=0.1$ or $\delta_0=-0.15$, as in the bottom row in Figure~\ref{fig:1Ddiffdelta}, then we find that $w=-1$ is excluded at 95\% confidence level (c.l.) in both cases: $w<-1.03$ and $w>-0.98$, respectively. However, these particular models are only included at  99.7\% c.l.

\begin{figure*}
\begin{center}
\includegraphics[width= .47\textwidth]{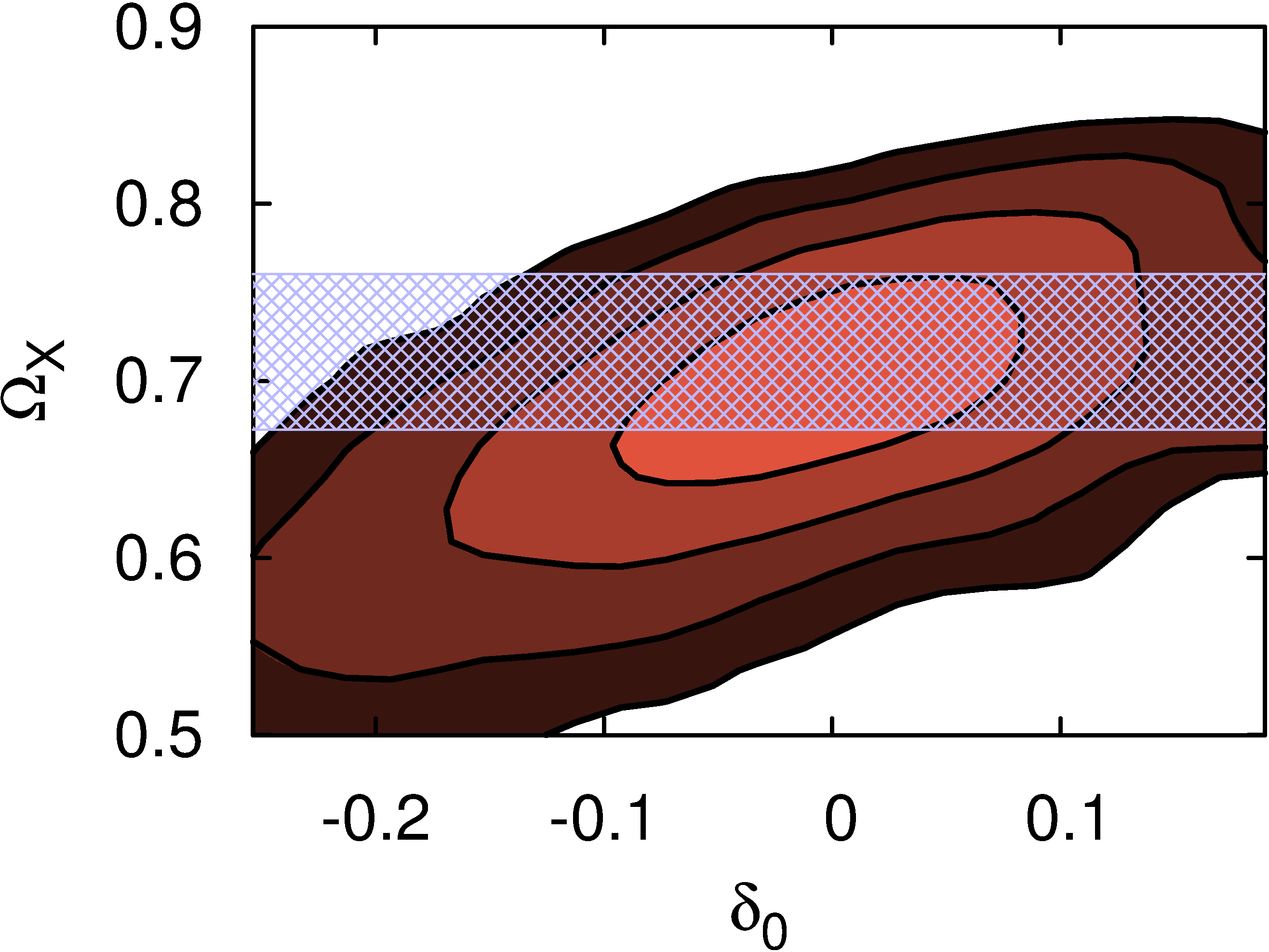}
\qquad
\includegraphics[width= .47\textwidth]{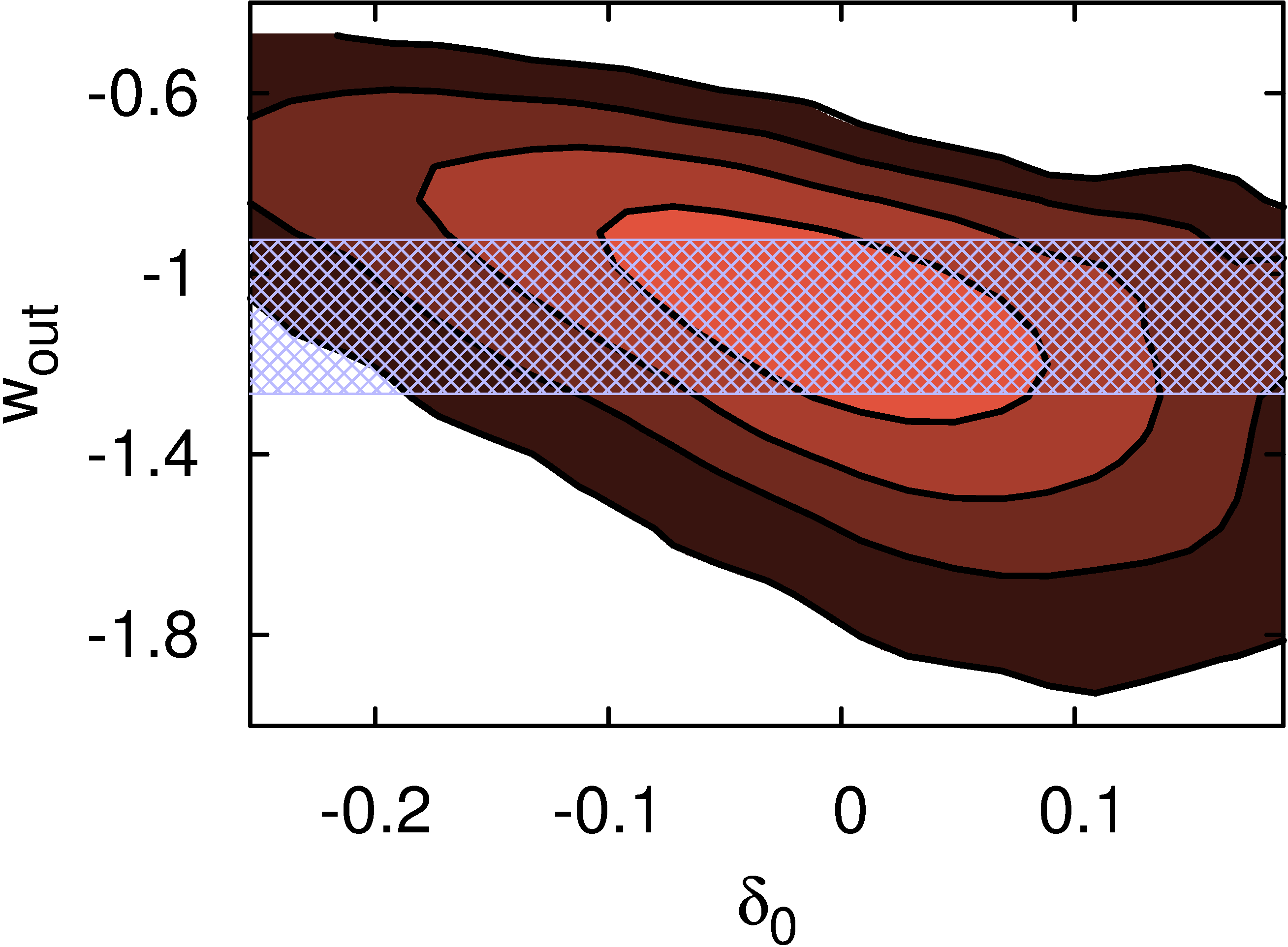}
\caption{Two-dimensional marginalized posterior probabilities for $\Omega_X$ and $w_{\rm out}$ with $\delta_0$, constrained by CMB, SNe, and $\hloc$. The color shaded regions correspond from the innermost region to the outmost region to 68\% c.l., 95\% c.l., 99.7 \% c.l.~and 99.99\% c.l., respectively. The blue horizontal band corresponds to the 95\% c.l.~one-dimensional constraints on $\Omega_X$ and $w_{\rm out}$ for the standard $w$CDM model.
This plot shows that if future data will constrain $\left| \delta_0 \right|$ to be large, then the inclusion of such data will shift the best fit region towards values of $w$ that are far from $-1$.}
\label{fig:degendirs}
\end{center}
\end{figure*}

In order to better show the degeneracy of $\Omega_X$ and $w_{\rm out}$ with $\delta_0$ we plot again in Figure~\ref{fig:degendirs} the corresponding two-dimensional marginalized posterior probabilities with 68\%, 95\%, 99.7 \% and 99.99\% confidence level contours. Also plotted for comparison are the 95\% c.l.~one-dimensional constraints on $\Omega_X$ and $w_{\rm out}$ for the standard $w$CDM model.
This plot is meant to justify the values $\delta_0=0.1$ and $\delta_0=-0.15$ used in the bottom row of Figure~\ref{fig:1Ddiffdelta}.
It shows indeed that large values of $\left| \delta_0 \right|$ that can significantly bias $\Omega_X$ and $w_{\rm out}$ are still within the 99.7\% c.l.

\subsection{Sensitivity to local Hubble-rate constraints} 

\begin{figure}
\begin{center}
\includegraphics[width= \columnwidth]{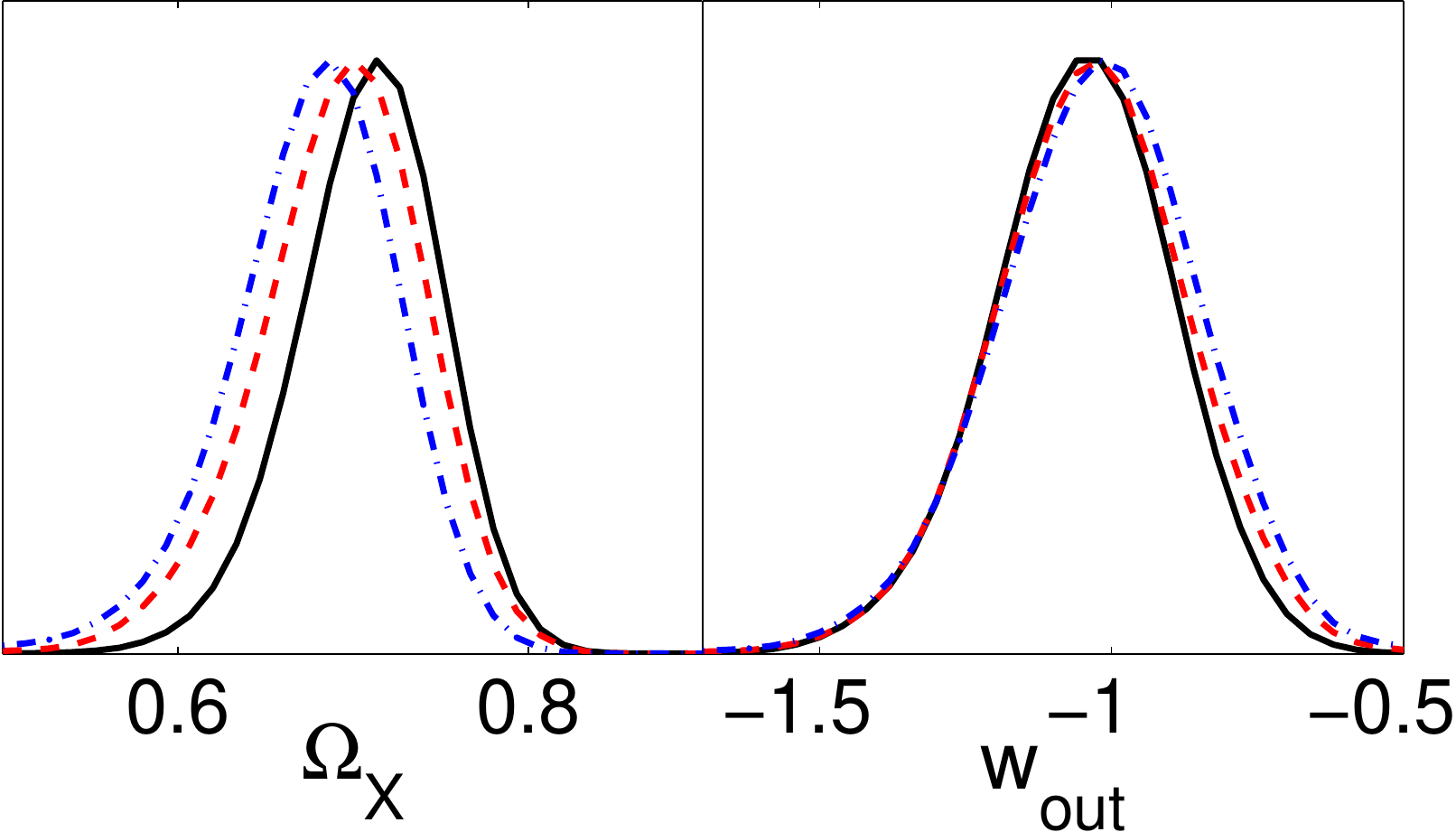}
\caption{One-dimensional marginalized posterior probabilities of $\Omega_X$ and $w_{\rm out}$ for the inhomogeneous model, given CMB, SN and $\hloc$ observations, comparing different  constraints on $\hloc$: \citet{Riess:2009pu} (solid black), \citet{Freedman:2000cf} (dashed red) and \citet{Sandage:2006cv} (dashed dotted blue).
The conclusions about the effect of the inhomogeneity on $w_{\rm out}$ are robust against different observational determinations of $\hloc$.}
\label{fig:H0dependence}
\end{center}
\end{figure}

In Figure~\ref{fig:H0dependence} we compare the effect of different Hubble-rate observations on the resulting posterior probabilities of $\Omega_X$ and $w_{\rm out}$, when we fit the $w$CDM model endowed with a local inhomogeneity to $\hloc$, SNe and CMB. As explained in Section~\ref{subsec:h0}, we compare the three different values from \citet{Riess:2009pu} (solid black), \citet{Freedman:2000cf} (dashed red) and \citet{Sandage:2006cv} (dashed dotted blue).
We see that the constraints on $\Omega_X$ do depend on the chosen measurement for $\hloc$, while the resulting constraints on $w_{\rm out}$ hardly depend on $\hloc$. This should be expected as SNe observations (constraining both $\Omega_X$ and $w_{\rm out}$) are insensitive to $\hloc$ while CMB observations (constraining $\Omega_X$ but weakly $w_{\rm out}$) are instead sensitive to $\hloc$.
These results imply that our conclusions with regard to $w_{\rm out}$ are robust against different observational determinations of~$\hloc$, and the constraints hence mostly follow from the SNe and~CMB.

\subsection{kSZ effect} \label{kSZe}

\begin{figure*}
\includegraphics[width= .44\textwidth]{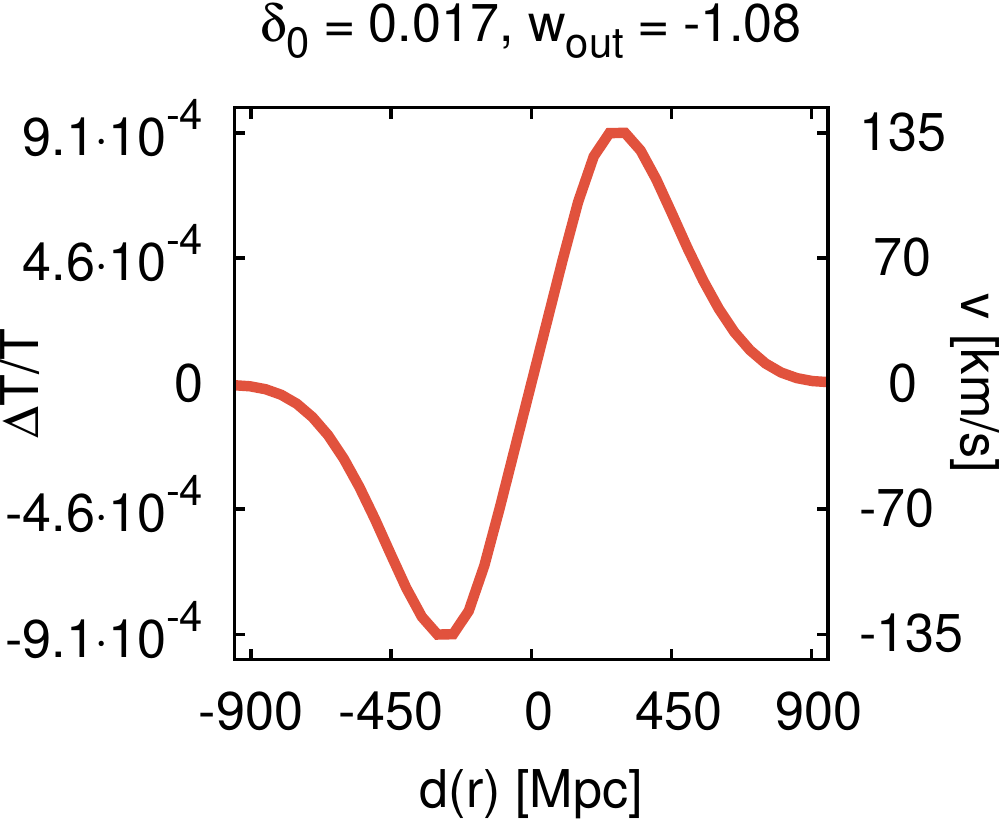}
\qquad \qquad \qquad
\includegraphics[width= .44\textwidth]{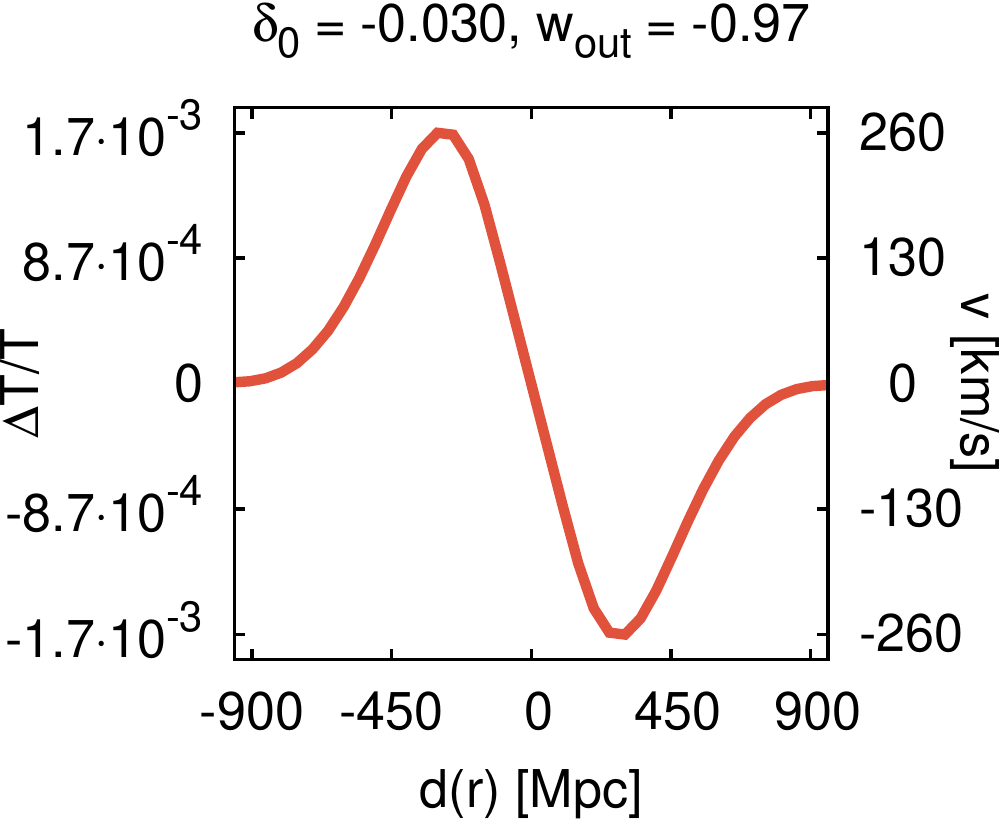}
\caption{The CMB-dipole observed by observers at different radii $d(r)$ in an LTB patch with a radius of 1 Gpc, for an over-density (left) and an under-density (right). On the right vertical axis we list the corresponding peculiar velocity that is derived assuming that the observed CMB dipole is caused solely by the peculiar velocity of the observer. The magnitude of the velocities does not exceed the magnitude of expected random velocities.
See Section \ref{kSZe} for more details.
}
\label{fig:dipoles}
\end{figure*}

In Figure~\ref{fig:dipoles} we show the CMB dipole that observers at different radii would observe, for a given configuration; an over-density on the left, an under-density on the right, in both cases the LTB patch has a radius of 1 Gpc, but with different values for $w_{\rm out}$. Both cases fit the data roughly as well as the standard $w$CDM, while still giving an interesting bias on $w_{\rm out}$.

We obtained these figures by -- at each radius -- starting an integration of the geodesic equations  in two directions (negative and positive $r$-direction), back to the surface of last scattering, which lies at constant time in the synchronous gauge of the LTB metric. The difference in redshift to the surface of last scattering in both directions is then translated into a $\Delta T _{\rm CMB}/ T _{\rm CMB}=(z_{+}-z_{-}) / (2+z_{ +}+z_{-}) $.
Only for small radii this is to a good approximation equal to the dipole in a spherical harmonics expansion of the CMB temperature map. On the vertical axis on the righthand side we list the corresponding peculiar velocity that an observed temperature difference corresponds to, if it were the effect of peculiar velocity alone. For both panes, left and right, the peculiar velocities do not exceed the magnitude of expected random peculiar velocities. Therefore the kinematic Sunyaev-Zel'dovich effect that such velocities induce on CMB photons~\citep{GarciaBellido:2008gd, Zhang:2010fa, Moss:2011ze} should be at present undetectable.
So as to strengthen this claim it is useful to look at the findings of \citet{Valkenburg:2012td} where constraints on the $\Lambda$LTB model from kSZ observations have been computed.
While in the present paper dark energy is not the cosmological constant, the above analysis should give nevertheless an estimate of the kSZ signal.
Therefore, this suggests that structures with a contrast of roughly $\sim$0.1 extending
for a radius of 1-2 Gpc are not excluded by present observations.  A thorough
study of the kSZ effect in these models is left to future work.

\subsection{FLRW Observer's $w(z)$} \label{wofz}

\begin{figure*}
\includegraphics[width= .45\textwidth]{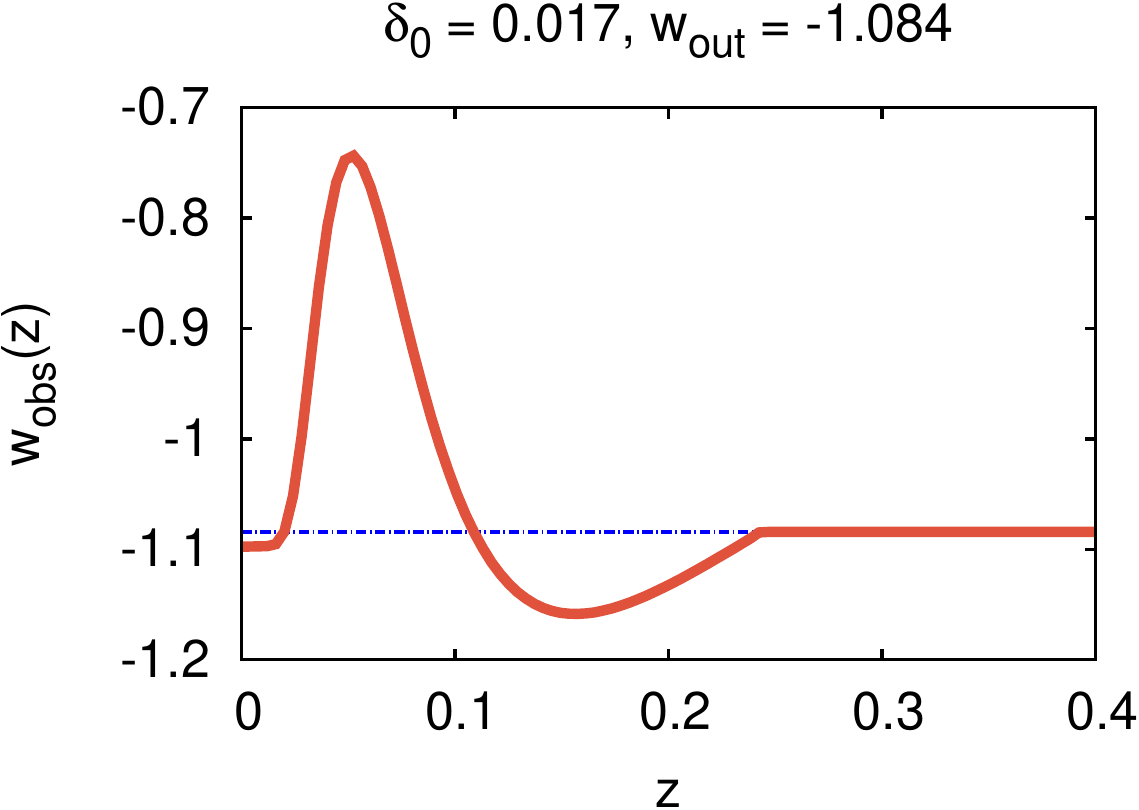}
\qquad
\includegraphics[width= .45\textwidth]{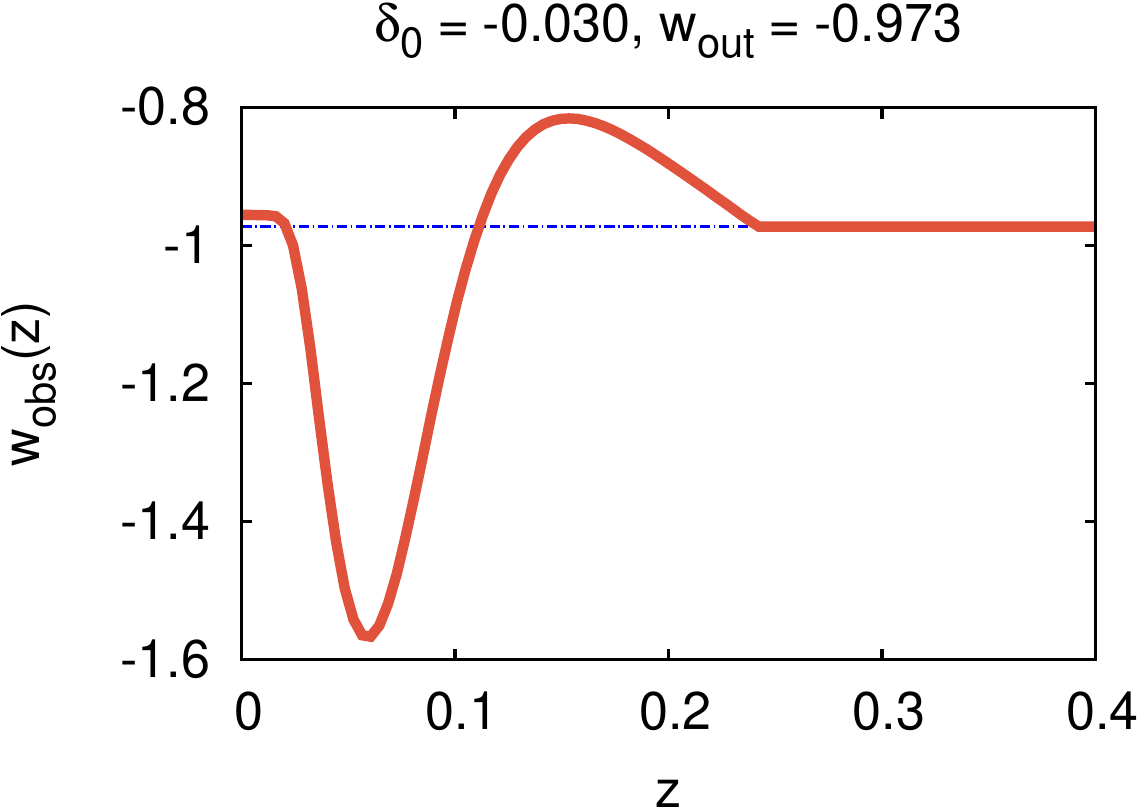}
\caption{The function $w_{\rm obs}(z)$ as defined in \citet{Clarkson:2007bc}, which is the equation of state of dark energy that an observer  thinks to see if the observer falsely assumes that the universe is described by the FLRW metric. The examples shown here are the same two cosmologies as in Fig~\ref{fig:dipoles}; an over-density (left) and an under-density (right). In both cases the observed $w_{\rm obs}(z)$ (solid red) matches closely with the fundamental $w_{\rm out}$ (dashed blue) once the radius is reached where the spherically symmetric metric matches to the surrounding FLRW metric. The inhomogeneity causes clear features in $w_{\rm obs}(z)$: the contracting core of the over-density increases the magnitude of $w_{\rm obs}(z)$, while the under-dense compensating shell has the opposite effect.  For the under-dense center, the inverse holds. In this picture $w_{\rm obs}(z)$ is computed using the exact solutions, while taking second derivatives of observed distances would not necessarily reveal these features.}
\label{fig:wofz}
\end{figure*}

Following \citet{Clarkson:2007bc} one can, given a luminosity distance-redshift relation in a homogeneous universe, compute what the underlying $w(z)$ of the dark-energy fluid is. In the homogeneous universe (described by the FLRW metric), one can find indeed an exact relation between $w(z)$ and the first and second derivatives of the luminosity distance with respect to redshift and two more parameters, $\Omega_k$ and $\Omega_{\rm m}$. If an observer knows the latter two parameters from other observations, and deduces  the first and second derivatives of the luminosity distance from SN observations, the observer can derive $w(z)$. In the scenario studied here, at background level $w$ is not a function of time or redshift. However, the inhomogeneity comes into play in the luminosity distance-redshift relation. Therefore, an observer that falsely assumes that the metric surrounding him/her is FLRW will in fact see a redshift dependence in $w$. We calculate the observed $w_{\rm obs}(z)$ (Eq. (3) in ~\citealt{Clarkson:2007bc}) for the two example models of Fig.~\ref{fig:dipoles}, and show the result in Fig.~\ref{fig:wofz}. Inside the inhomogeneity,  $w_{\rm obs}(z)$ shows a very clear signature of the matter distribution: the contracting core and expanding compensating shell for the over-density show corresponding effects on $w_{\rm obs}(z)$. The inverse holds for the under-density.

\begin{figure}
\begin{center}
\includegraphics[width= \columnwidth]{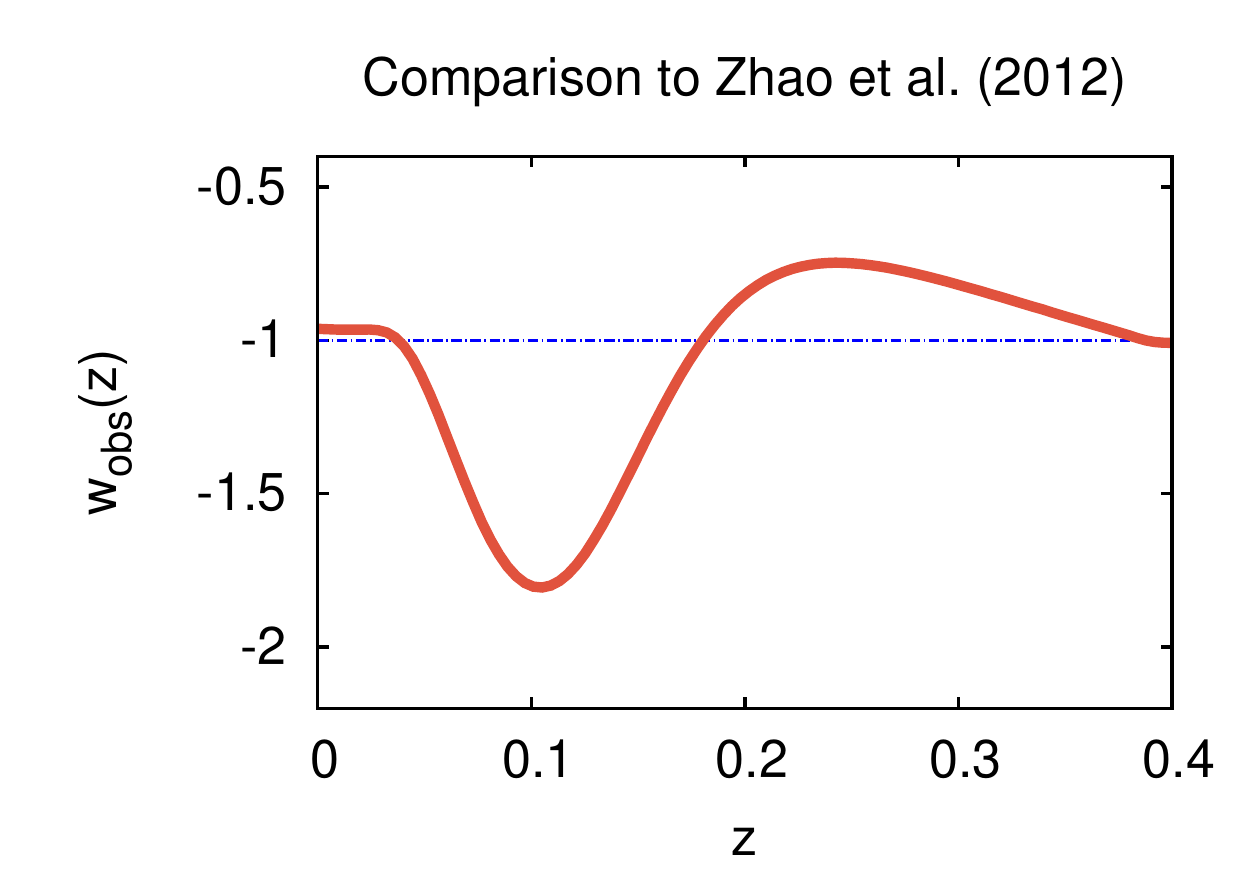}
\caption{As in Fig.~\ref{fig:wofz} but for a $\Lambda$CDM model endowed with a local underdensity of central contrast $\delta_{0}=-0.06$ and redshift boundary $z_{b}=0.4$ such that the observed $w_{\rm obs}(z)$ appears qualitatively similar to the $w(z)$ reconstruction by \citet{Zhao:2012aw} (see e.g. the pane (A2) of Fig.~1 in that reference).
If interpreted within this framework, the results of \citet{Zhao:2012aw} could indicate the presence of a large-scale underdensity around us, rather than a possibly time-dependent dark-energy equation of state.}
\label{fig:Void_wofz_zhao}
\end{center}
\end{figure}

Therefore, if one performs an analysis such as in \citet{Shafieloo:2009ti, Zhao:2012aw}, one may find a significant deviation from a constant $w$, while fundamentally $w$ is constant at the background level.
In particular, it is very interesting to note that the $w(z)$ reconstruction by \citealt{Zhao:2012aw} (see e.g. the pane (A2) of Fig.~1 in that reference), if interpreted within this framework, could indicate the presence of a large-scale underdensity around us, and not of a possibly time-dependent equation of state.
It is indeed worth noting the similarity of the result by \citet{Zhao:2012aw} with the observed $w_{\rm obs}(z)$ shown in Fig.~\ref{fig:Void_wofz_zhao}, which corresponds to a $\Lambda$CDM model endowed with a local underdensity of central contrast $\delta_{0}=-0.06$ and redshift boundary $z_{b}=0.4$.

\subsection{Other parameters} \label{showoff}

\begin{figure*}
\begin{center}
\includegraphics[width= \textwidth]{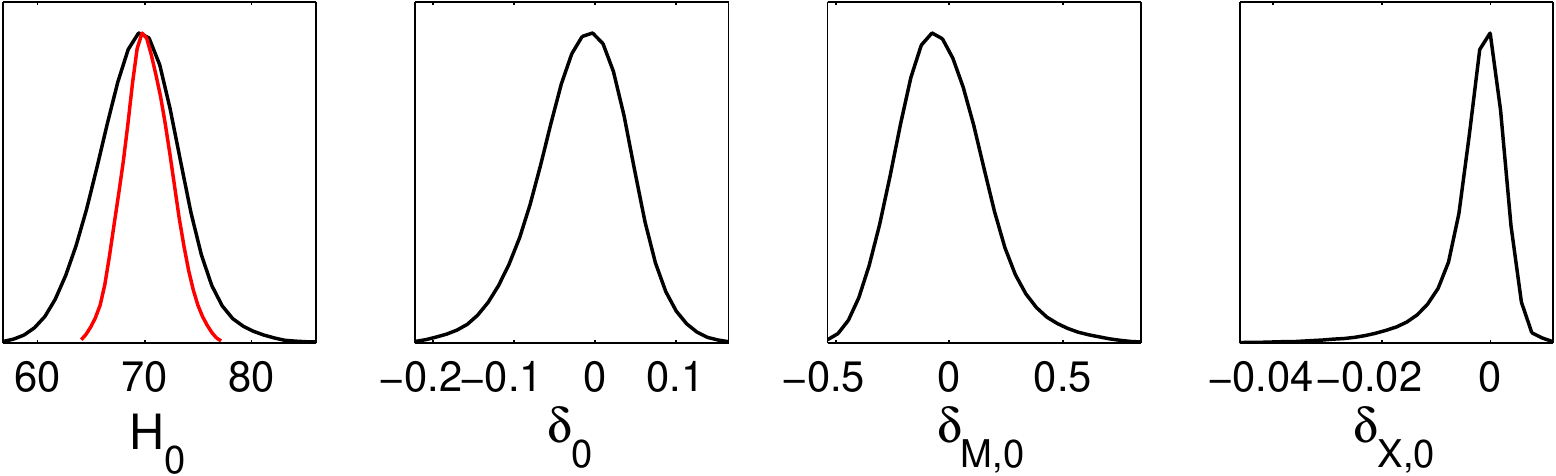}
\caption{One-dimensional marginalized posterior probabilities of $\hbkg$, which describes the age of the universe and does {\em not} correspond to the locally observed expansion rate, $\delta_0 \equiv \frac{\rho_{\rm in} - \rho_{\rm out}}{ \max(\rho_{\rm in},\rho_{\rm out})}$, $\delta_{M,0} \equiv \frac{\rho_{M,{\rm in}}}{\rho_{M,{\rm out}}}-1$, and $\delta_{X,0} \equiv \frac{\rho_{X,{\rm in}}}{\rho_{X,{\rm out}}}-1$. The presence of the spherical structure weakens significantly the bounds on the background expansion rate (in red we show the constraint on $\hbkg$ in the homogeneous $w$CDM model). Secondly, the local energy perturbation consists of an almost negligible dark-energy perturbation and a significant dust perturbation, which is not apparent if one considers the total $\delta_0$ alone. }
\label{fig:otherpars}
\end{center}
\end{figure*}

In Figure~\ref{fig:otherpars} we show posterior probabilities of $\hbkg$, which 
is the expansion rate of the background universe (and does {\em not} correspond to the locally observed expansion rate), $\delta_0 \equiv \frac{\rho_{\rm in} - \rho_{\rm out}}{ \max(\rho_{\rm in},\rho_{\rm out})}$, $\delta_{M,0} \equiv \frac{\rho_{M,{\rm in}}}{\rho_{M,{\rm out}}}-1$, and $\delta_{X,0} \equiv \frac{\rho_{X,{\rm in}}}{\rho_{X,{\rm out}}}-1$. All other parameters that we allowed to vary, listed in Table~\ref{tab:priors}, show practically no deviation in their constraints in the presence and absence of the inhomogeneity. The constraint on $\hbkg$ is significantly weakened when one takes into account the possibility that we may live in a local inhomogeneity. This result is similar to the findings in \citet{Valkenburg:2012ds}, even if the local inhomogeneity considered is orders of magnitude different.
Moreover, the actual matter perturbation that is allowed by the data can be as large as $\left|\delta_{M, 0}\right|\simeq 0.5$, leaving the total density perturbation around $\left|\delta_0\right|\simeq0.1$, because the energy perturbation in the dark-energy fluid is generally small as $w_{\rm out}$ is never very far from -1.

\subsection{Probability under homogeneous initial conditions} \label{gaussianperv}

Given the fact that we observe several large spots in the CMB, and the possibility that these spots are the result of density perturbations on the surface of last scattering, we can argue that living in such a perturbation must have a non-zero probability.\footnote{We would like to point out that for the almost-linear models considered in this paper the observer does not need  be very close to the center so as not to see a too large CMB dipole. For the cases of Fig.~\ref{fig:dipoles}, for example, the dipole is never larger than the observed value of $\sim 10^{-3}$. For larger contrasts there will be regions where observers would see a larger-than-measured dipole, but these regions will not occupy the majority of the inhomogeneity. Finally, we would like to stress that we have placed the observer at the center simply to simplify the numerical calculations.} If all perturbations arise from a smooth, close to scale invariant spectrum of primordial perturbations, it is not obvious that large cold and hot spots should exist, and there is an ongoing debate about this topic \citep{Cruz:2006sv,Zhang:2009qg,Ayaita:2009xm}.

Let us nonetheless quantify the probability of having the density perturbations that we took as examples for Figs.~\ref{fig:dipoles}~and~\ref{fig:wofz}. If these perturbations come from the same spectrum as the perturbations that we observe in the CMB, then the probability of their existence can be approximated by the variance of the gaussian density field, smoothed by a top hat filter with a radius that corresponds to the radius of the density perturbation under consideration \citep{Kolb:1990vq}. It must be noted that, because of the compensated shape of the density profile that we consider, taking the full radius of the spherical patch, $r_b$, as the radius of the top hat filter would give almost zero density perturbation. Therefore we choose the radius at which the density changes sign as the smoothing radius. This is roughly at $r_b/2$, but we use the numerically obtained exact value. Comparing at the time of decoupling when dark energy is negligible, and with the primordial spectrum of perturbations of the two models respectively (since the models were fit to the CMB, they carry spectral parameters), we find that the over-dense model considered in Figs.~\ref{fig:dipoles}~and~\ref{fig:wofz} is at three times the dispersion of the smoothed density field of its cosmology (CMB spectrum), and the under-dense model is at six times the dispersion of its cosmology. Notably the over-dense model is not at all unlikely to occur, while it does give a large effect on both $w_{\rm out}$ and $w_{\rm obs}$ as shown in Fig.~\ref{fig:wofz}.

\section{Conclusion} \label{conclusions}

We have analyzed present observations of the local expansion rate, distant supernovae and the cosmic microwave background within a flat $w$CDM model endowed with a local almost-linear inhomogeneity surrounding the observer. We have found a significant impact on the dark-energy parameters, in particular on the equation of state which is strongly degenerate with the inhomogeneity contrast. The implications of this degeneracy are twofold. On one hand we have shown that with prior knowledge on the inhomogeneity, to be obtained possibly with some future probe, it is already possible to rule out the case of the cosmological constant with current data.
On the other hand, even if future probes exclude the case of the cosmological constant in a homogeneous universe, this still may be due to a poor modeling of the large-scale structure of the universe.
The same conclusions apply to constraints on the time variation of the equation of state.

The analysis in the present paper is but a first step towards a more accurate reconstruction of the cosmological parameters.
We have indeed chosen, for technical reasons, a very specific dark-energy model and inhomogeneity profile. Before drawing definitive conclusions, a more comprehensive analysis should be performed.
Firstly, it would be particularly interesting (even though perhaps challenging) to consider a nonzero dark-energy sound speed. However, since we found that the dark-energy component is only very mildly inhomogeneous, we do not expect our results to be strongly dependent on the assumption of a negligible sound speed.
Secondly, it would be interesting to consider more inhomogeneous patches with more general density profiles, possibly with the observer at randomized positions \citep{Marra:2007pm, Marra:2007gc, Valkenburg:2009iw,Szybka:2010ky,Flanagan:2011tr}.

\section*{Acknowledgments}

It is a pleasure to thank Luca Amendola and Ignacy Sawicki for useful comments and discussions.
MP acknowledges financial support from the Magnus Ehrnrooth Foundation. 
VM and WV acknowledge funding from DFG through the project TRR33 ``The Dark Universe''.

\bibliographystyle{mn2e_eprint}
\bibliography{wbias}

\appendix

\section{Effect of constraints on priors} \label{prioco}

\begin{figure*}
\begin{center}
\includegraphics[width=0.8\textwidth]{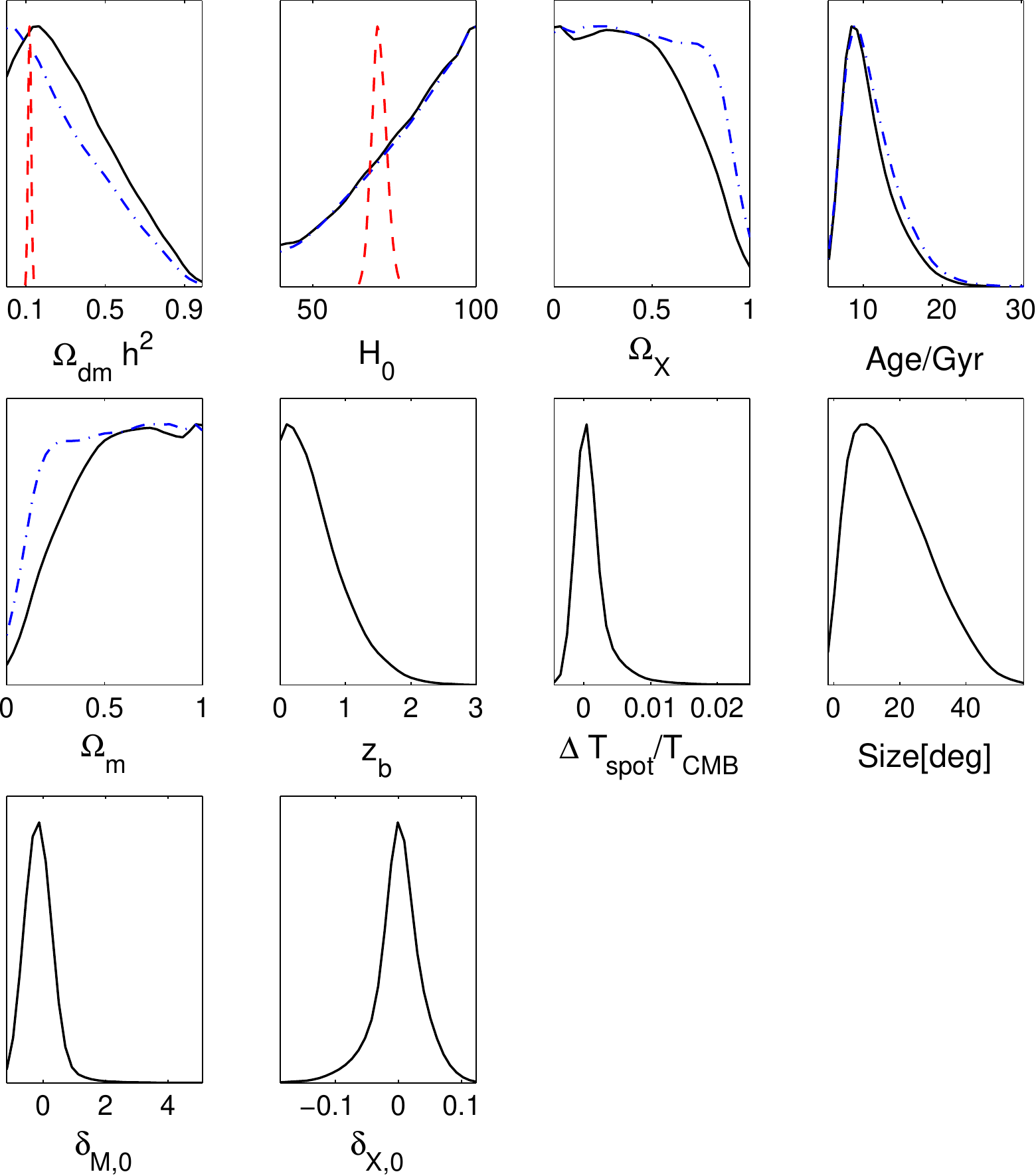}
\end{center}
\caption{Priors on $\Omega_{\rm dm}h^2$, $\hbkg$ and derived parameters, displaying those that are non-flat. The black (solid) lines represent the priors when the inhomogeneous metric is used, while the blue  (dash-dot) lines represent the priors in the homogeneous $w$CDM case. For comparison, in red (dashed) we show the constraints on $\Omega_{\rm dm} h^2$ and $\hbkg$ when the homogeneous $w$CDM case is fit to the data. The narrowness of the red curves shows that the data are very constraining on these parameters, and so the non-flat prior is of no importance. These prior distributions are obtained by running the MCMC analysis without any data, that is, accepting all points in parameter space with equal probability.}\label{fig:priors}
\end{figure*}

In Figure~\ref{fig:priors} we show the prior probability of $\Omega_{\rm dm}h^2$, $\hbkg$ and derived parameters, since they dependent non-linearly on combinations of input parameters on which we take flat priors, and hence their prior probability is non-flat. We obtained these prior probabilities by running the MCMC analysis without any data, accepting all points in parameter space with equal probability. Maybe surprisingly, the actual prior probabilities on $\Omega_{\rm dm} h^2$ and $\hbkg$ are not flat, even though we do list their prior ranges in Table~\ref{tab:priors} as flat, and we indeed gave flat priors on these parameters in the input of the MCMC simulation. The non-flatness stems from the additional constraint $\Omega_{X} >0$, demanding that the energy density of the dark-energy fluid is positive (if the Dark Energy is a pure cosmoillogical constant nothing prevents $\Omega_\Lambda<0$ from happening). Because we set $\Omega_k=0$, this condition is satisfied only when $\Omega_{\rm m}<1$, or $\Omega_{\rm m}h^2<h^2$.

To explain the relation, let us simplify the priors to $0<\Omega_{\rm m}h^2<1$ and $0<h<1$. The starting point is a flat prior on these parameters, {\em i.e.} that the probability $P(\Omega_{\rm m}h^2)=$ constant and $P(h)=$ constant, normalized such that $\int_0^1 P(x)dx=1$ for x being both $h$ and $\Omega_{\rm m}h^2$. If we write shorthand notation $C$ for the condition that $\Omega_X>0$ and $\Omega_k=0$, then imposing $C$ we have the conditional joint probability,
\begin{align}
P(\Omega_{\rm m}h^2, h ) \propto \left\{ \begin{array}{ll}P(\Omega_{\rm m}h^2, h | C)=\mbox{constant} &\mbox{if } C \mbox{ is true} \\ 0 &\mbox{if }  C \mbox{ is false} \end{array}\right. , 
\end{align}
normalized such that
\begin{equation}
\int_0^1\int_0^{1} P(\Omega_{\rm m}h^2, h | C)\,  \textrm{d}(\Omega_{\rm m}h^2) \, \textrm{d}h=1 \, .
\end{equation}
Then we find for the probabilities of $h$ and $\Omega_{\rm dm} h^2$, up to normalization constants,
\begin{align}
P(h|C)&= \int_0^{1}P(\Omega_{\rm m}h^2, h | C) \textrm{d}(\Omega_{\rm m}h^2) \nonumber\\
&\propto  \int_0^{h^2} \textrm{d}(\Omega_{\rm m}h^2) \propto h^2 \,,  \\
P(\Omega_{\rm m}h^2|C)&=\int_0^{1} P({\Omega_{\rm m}h^2}, h | C) \, \textrm{d}h \propto\int_{h=\sqrt{\Omega_{\rm m}h^2}}^1\, \textrm{d}h \nonumber \\
&\propto1-\sqrt{\Omega_{\rm m}h^2} \,,
\end{align}
in agreement with the priors in Figure~\ref{fig:priors}, favouring large $h$ and small $\Omega_{\rm m} h^2$.
However, as can be seen from the red dashed curves in Figure~\ref{fig:priors}, the data are constraining $\Omega_{\rm m}h^2$ and $\hbkg$ so tightly, that the non-flatness of the prior has no effect on the final parameter estimation.

\bsp

\label{lastpage}

\end{document}